%% file: 00-main.tex
\documentclass[10pt,twocolumn]{article}

\usepackage[letterpaper,margin=0.75in]{geometry}
\usepackage{amsmath,amssymb,amsfonts}
\usepackage{graphicx}
\usepackage{subfig}
\usepackage{cite}
\usepackage{algorithmic}
\usepackage{textcomp}
\usepackage{xcolor}

\title{
\textbf{EXR: An Interactive Immersive EHR Visualization in Extended Reality} \\
\vspace{0.5em}
\large\textit{Preprint version. This paper has been accepted to IEEE ICIR 2025. 
This is the author-prepared version and not the final published IEEE version. 
The final version will appear in IEEE Xplore.}
}

\author{
\begin{tabular}{c}
Benoit Marteau$^{1}$*, Shaun Q. Y. Tan$^{1}$*, Jieru Li$^{1}$, Andrew Hornback$^{2}$,\\
Yishan Zhong$^{1}$, Shaunna Wang$^{3}$, Christian Lowson$^{4}$, Jason Woloff$^{4}$,\\
Joshua M. Pahys$^{5}$, Steven W. Hwang$^{5}$, Coleman Hilton$^{4}$, May D. Wang$^{1,3}$ \\
\small
$^{1}$School of Electrical and Computer Engineering, Georgia Institute of Technology \\
$^{2}$School of Computer Science, Georgia Institute of Technology \\
$^{3}$Wallace H. Coulter Department of Biomedical Engineering, Georgia Tech \& Emory University \\
$^{4}$Shriners Hospitals for Children, Headquarters, Tampa, FL \\
$^{5}$Shriners Hospitals for Children, Philadelphia, PA \\
\small *These authors contributed equally.\\
\small Correspondence: stan99@gatech.edu, bmarteau3@gatech.edu
\end{tabular}
}

\date{}

\begin{document}
\maketitle

% ---------- Abstract ----------
\begin{abstract}
This paper presents the design and implementation of an Extended Reality (XR) platform for immersive, interactive visualization of Electronic Health Records (EHRs). The system extends beyond conventional 2D interfaces by visualizing both structured and unstructured patient data into a shared 3D environment, enabling intuitive exploration and real-time collaboration. The modular infrastructure integrates FHIR-based EHR data with volumetric medical imaging and AI-generated segmentation, ensuring interoperability with modern healthcare systems. The platform’s capabilities are demonstrated using synthetic EHR datasets, and computed tomography (CT)-derived spine models processed through an AI-powered segmentation pipeline. This work suggests that such integrated XR solutions could form the foundation for next-generation clinical decision-support tools, where advanced data infrastructures are directly accessible in an interactive and spatially rich environment.
\end{abstract}


\vspace{0.5em}

% ---------- Main Content ----------
\input{01-introduction}

\input{02-background}
\input{03-methodology}
\input{04-results}
\input{05-discussion}
\input{06-conclusion}

% ---------- Bibliography ----------
\bibliographystyle{unsrt}
\bibliography{Bibliography}

\end{document}

%% file: 01-introduction.tex
\section{Introduction}
Major advances in implementation science and data infrastructure within healthcare led to massive modernization of healthcare Artificial Intelligence (AI) infrastructures. State-of-the-arts data standards, such as the Observational Medical Outcomes Partnership Common Data Model (OMOP CDM)\footnote{https://www.ohdsi.org/data-standardization/}, the HealthLevel Seven (HL7) Fast Healthcare Interoperability Resources (FHIR)\footnote{https://fhir.org/}, or Digital Imaging and Communications in Medicine (DICOM) have enabled interoperable large-scale data infrastructures. These standards driving Electronic Health Records (EHRs) are a cornerstone of modern healthcare systems, enabling clinicians to make informed, evidence-based decisions\cite{tang2024harnessing, tong2023integrating, prabhu2022multi, shi2024ehragent}. These records contain a variety of patient information, including demographics, medical history, lab test results, and imaging data. This wealth of data forms the foundation of EHR systems, supporting clinicians in diagnosing, monitoring, and managing patients over time \cite{Isgut2024}. In addition to aiding clinical decisions, the extensive EHR data available has empowered researchers to develop advanced predictive models and assess various medical conditions more effectively, leading to better healthcare outcomes \cite{yuan2021performance, nickson2023prediction, Hornback2024}.

However, the sheer volume and complexity of EHR data can present significant challenges for healthcare professionals \cite{mazur2019association, Hornback2022, asgari2024impact}. As patient records become increasingly detailed, clinicians are often faced with an overwhelming amount of information, which can hinder their ability to quickly and accurately assess a patient’s condition. One major limitation is the use of traditional 2D interfaces for displaying longitudinal EHR data. On flat screens, clinicians must scroll through extensive lists of data, often leading to critical information being obscured or out of view \cite{khairat2021physician}. This format can hamper a holistic understanding of a patient’s health, especially when considering the temporal aspects of medical history or making cross-references between different types of data. Additionally, some medical data, such as DICOM images and volumetric 3D models, cannot be intuitively displayed on a 2D screen, making it harder for clinicians to interpret complex anatomical information. These limitations highlight the need for innovative visualization techniques that can better represent the full scope of EHR data in a more integrated and accessible format.

Extended Reality (XR) offers a promising solution to this problem. Unlike 2D screens, XR enables immersive, spatial interaction with medical data, allowing clinicians to visualize patient information in three dimensions. By leveraging XR, clinicians can interact with 3D anatomical models, medical images, and longitudinal data in an intuitive, real-time environment \cite{popescu2021dicom, marteau2024brain}. One key advantage of XR is its ability to display data in a more contextualized and holistic manner. For example, EHR data, including timelines of past medical conditions and lab tests, can be displayed alongside corresponding medical images (such as DICOM slices or 3D models) in a spatial arrangement. This immersive visualization enables clinicians to track the progression of diseases or conditions over time, facilitating longitudinal assessments in a way that is not possible with traditional 2D displays.

Furthermore, XR provides a level of interaction that traditional interfaces cannot match. Clinicians can zoom in on specific areas of interest, rotate 3D models for a better view, and even simulate medical procedures or anatomical changes. This capability allows for a deeper understanding of complex structures, such as the brain or spine, which are difficult to interpret through conventional imaging methods. Moreover, the visualization and interaction can be collaborative, where the same shared virtual space enables clinicians to visualize data modalities side by side, facilitating decision-making and improving clinical workflows \cite{fischer2020volumetric}.

Despite the growing interest in XR for medical imaging, existing applications \cite{popescu2021dicom, pires2021use, marteau2024brain, ells2025free} typically focus on visualizing 3D DICOM images in isolation, without integrating these images with other critical patient data. In most cases, medical images are treated as separate entities, disconnected from the patient's full medical history. This approach does not reflect real-world clinical workflows, where clinicians must consider imaging data in conjunction with a patient's longitudinal health record \cite{kline2022multimodal, mohsen2022artificial}. To address this gap, our work introduces a multimodal XR infrastructure to visualize EHR data with other data modality (e.g., 3D imaging), integrating AI models to further extend visualization capabilities, specficially automated 3D spine segmentation AI model as a proof-of-concept. 

This work introduces a novel yet-to-be-validated framework for multimodal, FHIR-integrated XR visualization. While this study focuses on demonstrating architectural feasibility, upcoming efforts will emphasize empirical validation in real clinical settings.

The contributions of this work are:
\begin{itemize}
    \item \textbf{EHR immersive and interactive visualization within XR enviroment} integrating FHIR interoperability standard. 
    \item \textbf{Multimodal XR framework} unifying structured EHR data and unstructured volumetric imaging study and AI-generated 3D segmentations in collaborative workspace
    \item \textbf{Demonstrative use-case} showcasing the interactive visualization of EHR data and AI-assisted 3D spine segmentation, illustrating the potential for XR as a foundation for complex multimodal interactive tool.  
\end{itemize}

%% file: 02-background.tex
\section{Related Works}
\subsubsection{Multimodal EHR Visualization}
Visualization of clinical data has been a longstanding area of research, with efforts focusing on improving usability and cognitive load reduction for healthcare professionals. Traditional EHR interfaces have often been criticized for their fragmented presentation and lack of context awareness, which can lead to clinician burnout and errors. In response, interactive dashboards and timeline-based visualizations have been developed to better organize patient data and support temporal reasoning \cite{Zia2024}.

Advancements in medical imaging visualization have progressed from 2D slice-based viewers to volumetric rendering techniques that allow more detailed inspection of anatomical structures \cite{Wan2012}. These techniques have been incorporated into surgical navigation systems and diagnostic tools, providing surgeons and radiologists with enhanced spatial understanding. However, many existing systems rely on desktop-bound workstations, limiting flexibility and collaborative potential.

XR technologies have increasingly been explored for medical applications, offering immersive environments where clinicians can interact with complex data more intuitively \cite{Zhong2022}. Several studies have demonstrated the effectiveness of XR in surgical planning and training, allowing users to manipulate 3D models with natural gestures and explore anatomical structures at varying scales \cite{Ntakakis2023}. Beyond imaging, XR has been used to visualize physiological data and simulations, enhancing clinicians’ comprehension of patient conditions.

The integration of multimodal medical data into XR environments remains an emerging research area \cite{mti8110098}. Efforts to combine imaging data with clinical notes, lab results, and genomic information are still preliminary, often limited by interoperability challenges and the lack of standardized data fusion frameworks. Additionally, the application of AI within XR for automated data segmentation and anomaly detection is gaining traction, promising to augment clinical workflows by highlighting relevant features dynamically.

Collaboration in XR medical environments has also been studied, with multi-user platforms enabling remote consultation and education. These systems facilitate shared visualization and interaction with patient data, supporting multidisciplinary team decision-making \cite{Tan2024}. Despite these advances, fully integrated XR systems that unify comprehensive EHR data, medical imaging, and AI-powered analytics for real-time clinical use remain scarce, motivating further development in this domain.

\subsubsection{AI Integration: 3D Spine Segmentation}
In spinal image segmentation, U-Net \cite{ronneberger2015u} established the encoder–decoder paradigm for 2D semantic segmentation, with skip connections enabling accurate delineation from limited data. For 3D volumetric segmentation, V-Net \cite{milletari2016v} captures spatial context via volumetric convolutions, and nnU-Net \cite{isensee2021nnu} automates preprocessing, architecture tuning, and postprocessing to provide a robust cross-dataset baseline. Coarse-to-fine strategies are widely adopted: an initial stage localizes the region of interest or individual vertebrae, followed by fine segmentation. Examples include DenseUNet \cite{li2018h} and the SpatialConfiguration-Net (SCN) \cite{payer2019integrating} combined with U-Net \cite{payer2020coarse}, where SCN uses landmark detection and spatial configuration regression to incorporate global anatomical priors before 3D U-Net refinement, achieving strong results in the VerSe 2019/2020 challenge \cite{sekuboyina2021verse}.
\begin{figure*}[t]
    \centering
    \includegraphics[width=0.8\textwidth]{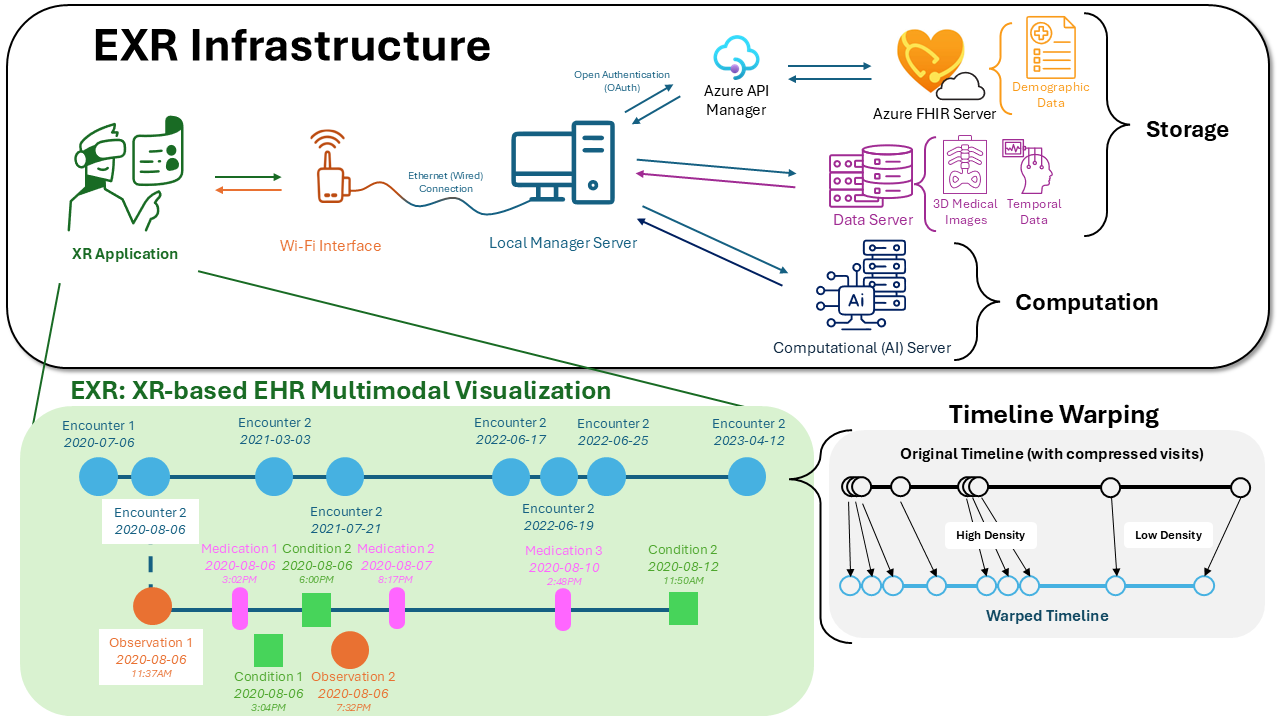}
        \label{fig:case1}
    
    \caption{EXR Infrastructure comprises of five components: (1) the XR Application, (2) the interface between the XR device and the local manager server, (3) a local manager server, (4) storage server (FHIR server and unstructure data storage), and (5) AI computational server. The EXR represents the different encounters on a warped timeline for a specific patient. For a specific encounter, we can represent the different events (e.g., conditions, observations, medications, etc.) as a detailed timeline under the main visit timeline. The timeline warping enables sparse and long timeline (over years) to not compress visits close to each other (over days)}
    \label{fig:EXR_overview}
    \vspace{-1em}
\end{figure*}
To overcome the limited receptive field of Convolutional Neural Networks (CNNs), Transformer-based models have been introduced. TransUNet \cite{chen2021transunet} integrates Vision Transformer (ViT) encoding with U-Net decoding to improve long-range dependency modeling, while UNETR (UNEt TRansformers) \cite{hatamizadeh2022unetr} uses a Transformer directly as a volumetric encoder with skip connections. Swin Transformers \cite{cao2022swin, hatamizadeh2021swin} and hybrid attention modules \cite{qiu2024agileformer, li2024verformer} have also demonstrated promising transferability. 
Instance-level segmentation extends beyond semantic masks to assign anatomical labels. Common strategies include detection-based pipelines with graph-based correction, iterative or memory-based Fully Convolutional Networks (FCNs) that sequentially segment vertebrae along the spine, and two-stage cascades with anatomical consistency constraints. Lessmann et al. introduced a memory-based sequential approach in CT \cite{lessmann2019iterative}, while Dense-U-Net and Mask R-CNN variants have been applied to vertebral and intervertebral segmentation \cite{cheng2021automatic}. In X-ray, VertXNet \cite{chen2024vertxnet} integrates U-Net and Mask R-CNN \cite{he2017mask} for vertebra detection and labeling. Meanwhile, the SPINEPS framework \cite{moller2025spineps} achieves both semantic and instance segmentation of the whole spine in T2w MRI through a two-phase pipeline.

%% file: 03-methodology.tex
\section{Methods}
\subsection{EXR: EHR XR Visualization Infrastructure}
Our infrastructure comprises of five components: (1) the XR application itself, (2) a network interface between the XR device and the local manager server, (3) a local manager server used to communicate with all the different elements in the infrastructure, (4) data storage, comprised of both FHIR server (e.g., Microsoft Azure FHIR) and unstructured storage server (e.g. local storage), and (5) AI computation. We repented the infrastructure in \textbf{Figure \ref{fig:EXR_overview}}.

We decided to adopt a modular approach that would also be compatible with existing healthcare systems. Specifically, each component in our infrastructure can be replaced depending on the needs and constraints of modern healthcare infrastructure. Indeed, healthcare environment presents great challenges regarding data transfer and visualization, as in the United States for example, patient data are considered Protected Health Information (PHI), with strong regulations surrounding the manipulation of such data. 

In effect, this means that the different components may realistically be connected to secured internal hospital networks. Therefore, we designed our infrastructure around a local manager server that would connect to both onsite or cloud storage and AI computational servers. Ideally, this local manager server would not be connected to the Internet, but instead connect to an internal secured network linked to data and computation. The advent of cloud services, such as Microsoft Azure, or Amazon Web Services (AWS) have enabled secured communication with scalable storage servers, introducing additional security challenges. To mitigate security threats, we incorporated the Microsoft Azure Open Authorization (OAuth) into our infrastructure, allowing granular data access to authorized users only. Although the EXR infrastructure operates through a local manager server instead of direct cloud connectivity, we do not believe that this design negatively impact dynamic data streaming. Conversely, the use of a local manager server can enhance data-stream stability, particularly in environments where the XR headset might otherwise depend on unreliable institutional wireless networks. By routing data through a local server (that can be either wired or wireless) within the same network, the system maintains a more consistent and controlled data flow.

To mitigate potential loss of semantic relationships due to modularity, our architecture leverages the FHIR data model as the semantic backbone. FHIR inherently represents the relationship between the different common data elements (FHIR resources such as  patients, encounters, observations, and procedures). By retaining these relationships within the local manager server, EXR preserves contextual links across data modules. Moreover, we developped both the local manager server and XR application in a modular manner, meaning that we can add add or remove  FHIR resources without compromising these semantic dependencies, notably because we plan to add more FHIR resources in future work.

We are now describing the different components of our infrastructure: the Azure FHIR Server which stored the EHR data, the local manager server, the XR application and interface with the local manager server, and the AI pipelines.

\subsubsection{Azure FHIR: EHR Data}
To develop and demonstrate the capabilities of our XR-based interactive EHR visualization, we deployed our own Azure FHIR server populated with Sythea synthetic data \cite{walonoski2018synthea}. We selected Synthea data to represent our EHR as it enabled us to have a safe development with data similar to what real-world healthcare systems encounter. We populated our FHIR server with more than 100 synthetic patients, containing patients demographic information, conditions (e.g., diagnoses), procedures, medications, encounters (e.g., visits), observations, imaging studies, immunization, diagnostic reports, and others.  

To add security and versatility to our infrastructure proof-of-concept, we adopted the Microsoft Azure OAuth flow, as well as an Azure Application Programming Interface (API) Manager (APIM), enabling user to authenticate with their Microsoft account, getting access only if authorized. This approach was explored by Marteau B. et al. when developing their Brain Digital Twin (DT) with FHIR data access \cite{marteau2024brain}. However, our Authentication mechanism is placed in the local manager server instead of the XR device itself to avoid potential compatibility issues with the authentication mechanism and devices (e.g., Android for our XR Device).

\subsubsection{Local Manager Server}
To enable data communication between the different components of our infrastructure, we developed our own Python-based Flask server \footnote{https://flask.palletsprojects.com/en/stable/}. The local manager server has three goals:
\begin{itemize}
    \item It handles authentication and communication to the Azure APIM and Azure FHIR server
    \item Transfer data to the AI model and manages it.
    \item Transfer data between the storage and FHIR server and the XR device.
\end{itemize}

Specifically, data are transferred as text, using JavaScript Object Notation (JSON), or byte streams. However, authorization tokens used to connect to the FHIR server are not shared with the XR device, reducing potential security risks. 

\subsubsection{XR Application and Interface with the Local Manager Server}
We developped our XR application on the Meta Quest 3 device using Unity (Unity 6000.1.15) as our game engine\footnote{https://unity.com/}. Specifically, we leveraged Meta XR All-in-One Software Development Kit (SDK) (Version 77.0) to enable faster development\footnote{https://developers.meta.com/horizon/downloads/package/meta-xr-sdk-all-in-one-upm/}. 

To enable the connection between the Meta Quest 3 and the local manager server, we designed our infrastructure with an interface component, which would handle the wired, Wi-Fi or Bluetooth connection to the XR device and the wired connection with the local manager server. The introduction of this interface allow the user to select which device could be connected to the local manager server, while avoiding the need to connect the XR device to the internet, notably as some hospital system may not allow direct connection of any "personal" device receiving patient data on their Wi-Fi network. In essence, this interface enabled the creation of a local network comprised of the local manager server and XR devices connected to it. Moreover, the user can code the interface to only accept connection to pre-approved devices (e.g., using their Media Access Control or MAC address whitelist). This also enables multiple XR device to connect to the same local manager server.

\subsubsection{AI Computational Server}
We designed our integration of AI pipeline as modular modules, where an AI pipeline is defined as a process by its input and outputs. We currently hard coded the different AI pipeline in the local manager server, althogh ideally this AI pipeline list could be dynamically updated when the user adds or removes AI models. 

we used Georgia Institute of Technology Partnership for an Advanced Computing Environment (PACE) computational server to store unstructured data and run the AI models for this proof-of-concept. However, in the case data and AI models are not in the same server, the local manager server should handle data transfer between these two components. 

\subsection{XR-based EHR data visualization}
We designed our interactive visualization to view three levels of EHR data: the first level representing the different patients in an EHR system. The first step is patient retrieval. We explored three approaches: the first approach was to group patients in a 3D graph, enabling the user to visualize patient clusters. While this approach enables a quick visualization of common phenotypes, a large number of patients can render patient retrieval impractical. A second approach was to develop a menu where the user can select phenotypes to retrieve a patient cohort. The last approach was to directly name the patient or retrieve a patient with her/his identification number. 
The second level of our visualization was to represent the different visits or encounters a patient had. However, this timeline generation presents a challenge: how to represent sparse visits that sometimes can span years. We carefully bound the timeline representation, warping the distance between two events. 

To properly represent sparse timeline, where visits can be days apart or years apart, we needed a warping mechanism capable of showing all visits in a reasonable manner (e.g., on a 2-meters wide line) representing years of visits without compressing close visits. To do so, we developped a warping formula that would switch between a linear representation for visits that are close to each other (e.g., few days) and have logarithm representation for visits far appart (e.g., years).
In this effect, we introduce this warping formula:
\begin{multline}
\Delta t = \left[1-\frac{\rho}{\rho_{max}}\right]\times e^{\frac{\rho}{\rho_{max}}} \times log(\Delta t) \\ + \frac{\rho}{\rho_{max}}\times e^{\left[1-\frac{\rho}{\rho_{max}}\right]} \times \Delta t
\end{multline}

Where $\rho$ represents the visit density, $\rho_{max} = max_{\rho}(\rho)$, and $\Delta t$ the difference between two visits. With this formula, when the visit density is high, the formula behaves linearly, and when the visit density is low, the formula behaves logarithmically. 
We tried different density formulas for our application, such as the number of visit per time period (e.g., (.e.g, $\frac{\# visits}{Time \; Window}$)), or the the inverse number of days between a visit and the previous one (.e.g, $\frac{1}{\#day \; since \; previous \; visit}$), or the inverse number of days between a visit and the next one (.e.g, $\frac{1}{\#day \; until \; next \; visit}$).
Ultimately, we selected the third option, with a visit density defined as the inverse number of days until next visit, as it provided the most intuitive timelines. We represented the timeline warping in \textbf{Figure \ref{fig:EXR_overview}}.

Lastly, the third level of representation relates to the different events (e.g., observations, procedures, conditions, etc.) during a visit. We represented the different type of data using different shapes (e.g., pill for medication, cube for procedure, sphere for observation, etc.). 
When clicking (or grabbing) one specific event (e.g., observation), we display certain information related to this resource. We also integrated 3D image visualization by modifying imaging studies present in our synthetic FHIR dataset to repents the location of generic spine data in the storage server. This enabled us to test our complete infrastructure, and enabled us to adapt and test the AI pipeline component. 
We represented the XR application timeline visualization in \textbf{Figure \ref{fig:EXR_overview}}

\subsection{Use-case: AI-based Vertebral Localization and Segmentation}
For our specific use-case, we adopted an existing AI model to perform 3D spine segmentation. This model comprises of two parts: a coarse-to-fine approach \cite{payer2020coarse} for vertebrae localization, labeling, and segmentation, described in previous work and evaluated on the VerSe benchmark \cite{sekuboyina2021verse}. This pipeline is performed in three stages. First, a low-resolution 3D U-Net predicts a heatmap of the vertebrae centerlines, whose centroids are used for global alignment and field-of-view cropping. A Spatial Configuration-Net then performs multi-point heatmap regression at a higher resolution, combining local appearance with global spatial configuration. Local maxima generate ordered vertebrae centers and labels from C1 to L5, while constraints on spacing and order maintain anatomical consistency, including variants such as T13 or L6. For segmentation, a single 3D U-Net operates on 1mm patches centered at each predicted point to produce binary masks. These predictions are mapped back to the original physical space and fused via voxel-wise maximum response to obtain a multi-label 3D mask. The output includes a mask and centroid for each vertebraenabling redering in the XR environment.

We adopted this approach because task decomposition reduces ambiguity between anatomical structures of similar shape. Direct multi-class segmentation often exchanges labels, while separating localization and labeling from the binary segmentation of each vertebra improves label consistency and boundary quality. Global alignment reduces domain shifts due to field of view changes, metal artifacts, and fractures. The spatial configuration module perceives long-range relationships of the spine and maintains stable predictions under partial views and anatomical variations. The model trained on VerSe 2020 (mean Dice score of 91.23\%) still maintains a mean Dice score of 86.44\% on the VerSe 2019 test set \cite{sekuboyina2021verse}, demonstrating robust generalization ability of this method. The output of this model conforms to our XR workflow, generating object-level masks and vertebral location labels, allowing for rapid segmentation of the spine without retraining.

To integrate this use-case to our proof-of-concept infrastructure, we manually modified imaging studies FHIR resources in our server to represent the original raw data used in this AI pipeline. We then run the whole AI models to localize and segment the different vertebrae of a spine, saved the results into the storage server and sent them to the XR application for visualization. 
\vspace{-1.5em}

%% file: 04-results.tex
\section{Results}
\subsection{EHR XR Visualization}
\begin{figure}[htbp]
    \centering
    \subfloat[]{%
        \includegraphics[width=0.9\linewidth]{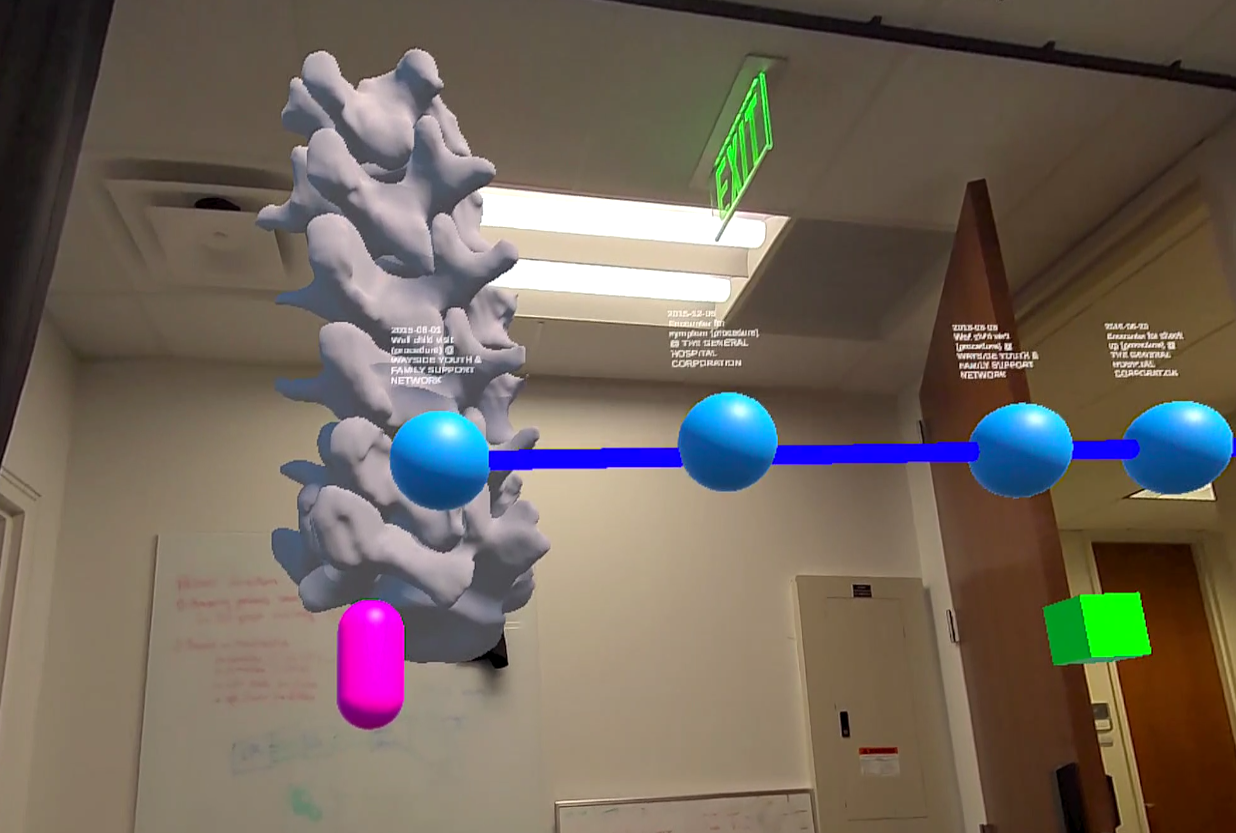}
        \label{fig:case1}
    }
    \caption{XR-based EHR visualization}
    \label{fig:EXR}
    \vspace{-1em}
\end{figure}

We represented in \textbf{Figure \ref{fig:EXR}} the current XR-based EHR visualization. We represented the different concepts with different shapes and colors. Moreover, we displayed useful information on top of each shape. Moreover, when clicking on a concept, the user can move it in space. Ultimately, when clicking on the imaging study (in pink in the figure), the system automatically loads relevant imaging and AI output (here, 3D segmentation). Text-based data, such as physician notes or diagnostic reports, are displayed as floating text panels adjacent to corresponding 3D icons. This contextual linking maintains readability while preserving spatial organization within the XR workspace.

While the current proof-of-concept employs distinct shapes and colors to differentiate between EHR resources (e.g., pills for medications, cubes for procedures, spheres for observations), we couldn't find a consensus regarding universally accepted and intuitive visual representations (icons) of these different events. Therefore, next steps of the project will help identify potential taxonomy. 

\subsection{Vertebral Localization and Segmentation}
We evaluated the model on the test sets from the VerSe2019 and VerSe2020 datasets. These test sets contain 113 subjects in total. Each CT volume was pre-processed by smoothing with a Gaussian kernel with $\sigma$ = 0.75 mm to reduce noise.

\begin{figure}[htbp]
\vspace{-1em}
    \centering
    \subfloat{%
        \includegraphics[width=0.25\linewidth]{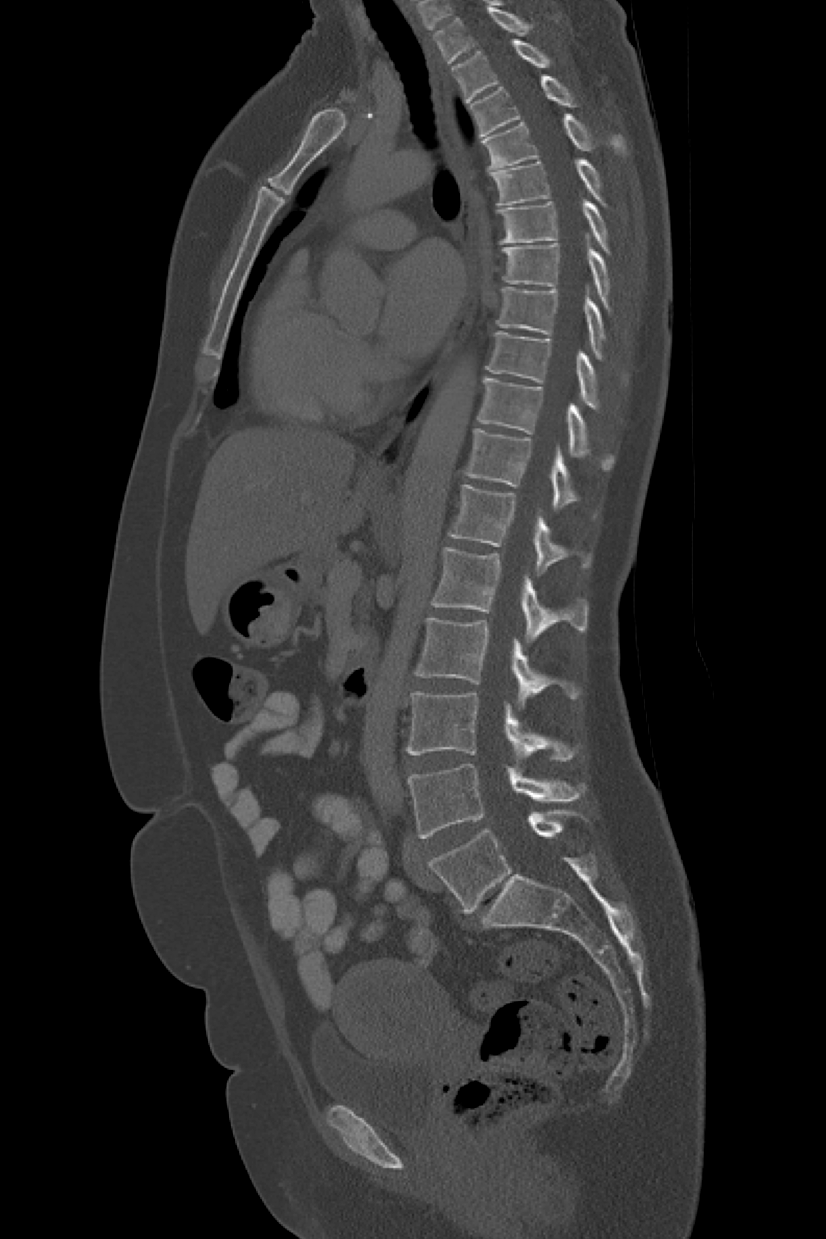}
        \label{fig:ct_image2}
    }
    \subfloat{%
        \includegraphics[width=0.25\linewidth]{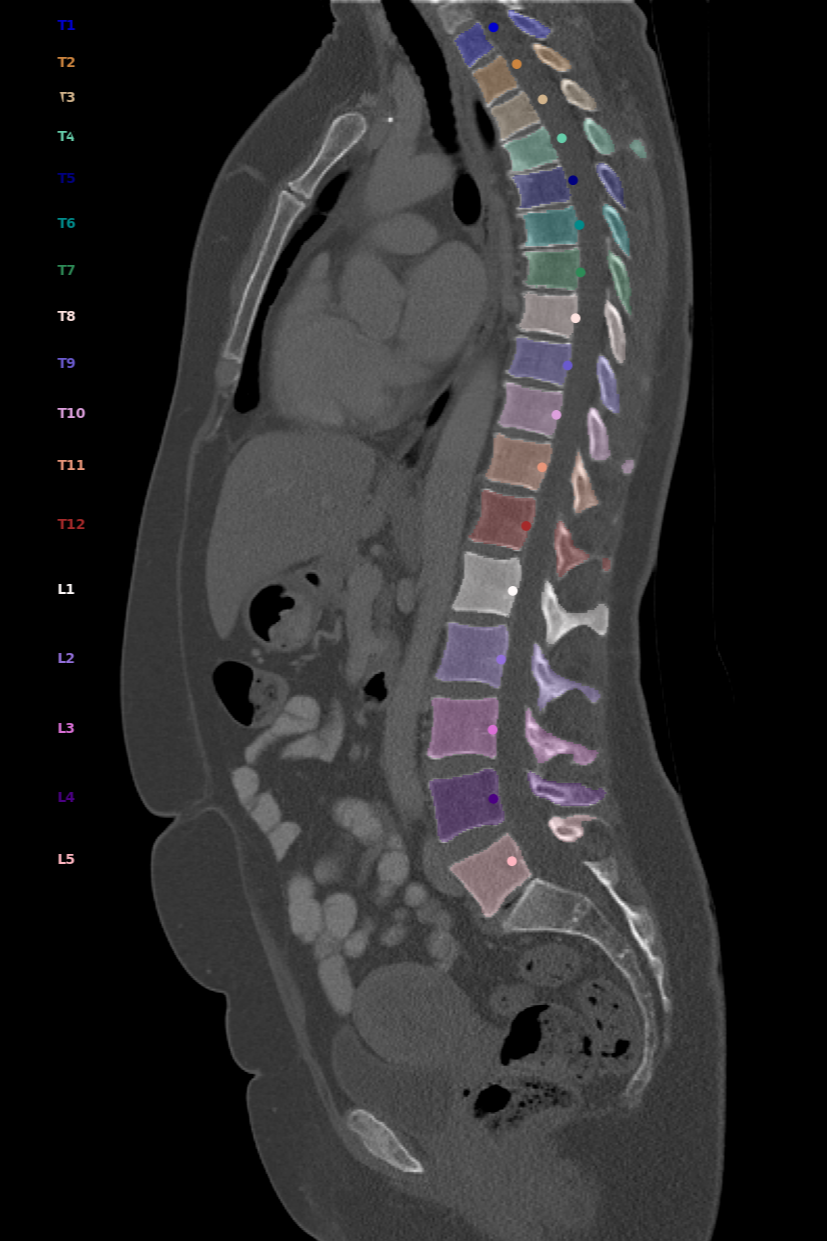}
        \label{fig:segmentation2}
    }
    \subfloat{%
        \includegraphics[width=0.25\linewidth]{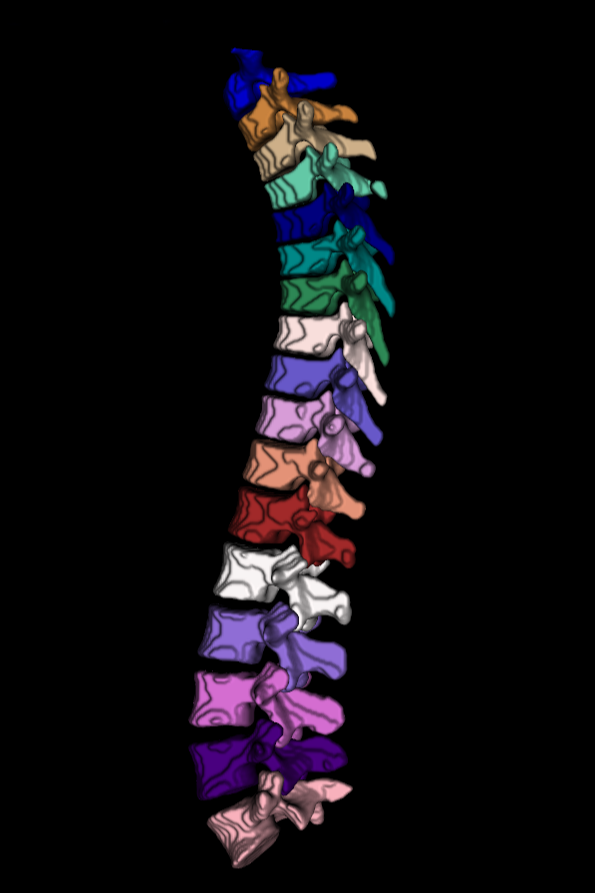}
        \label{fig:3d_mesh2}
    }
    \hfill
    \subfloat{%
        \includegraphics[width=0.25\linewidth]{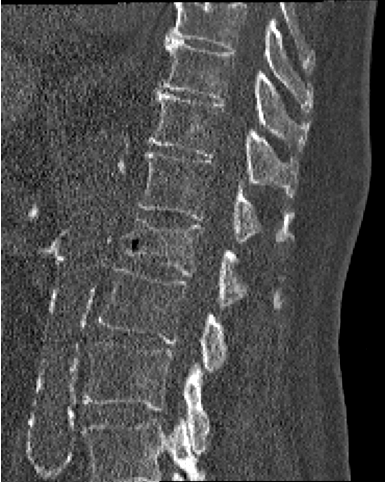}
        \label{fig:ct_image3}
    }
    \subfloat{%
        \includegraphics[width=0.25\linewidth]{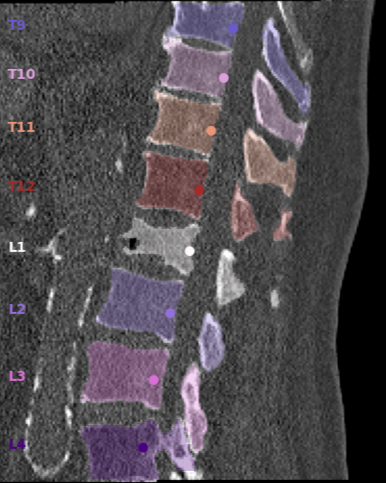}
        \label{fig:segmentation3}
    }
    \subfloat{%
        \includegraphics[width=0.25\linewidth]{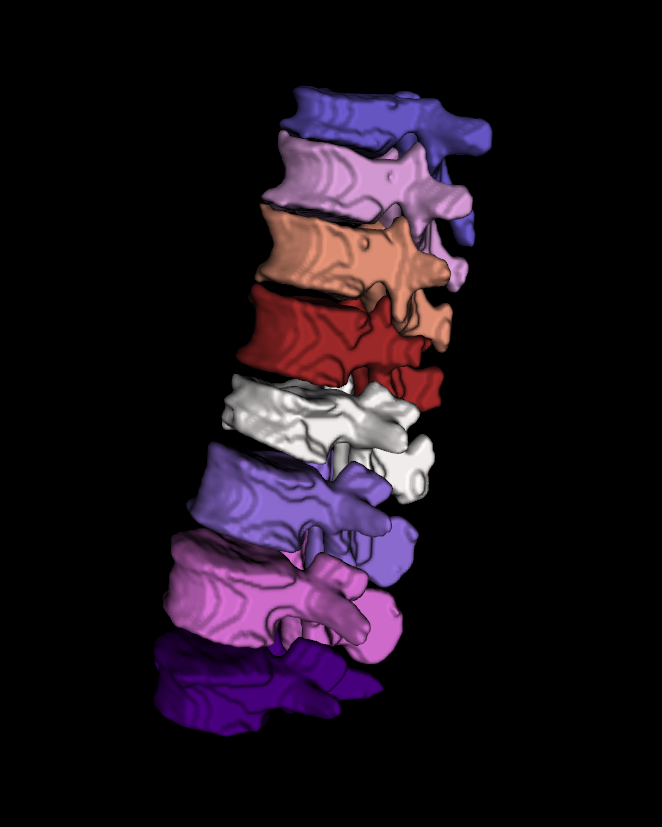}
        \label{fig:3d_mesh3}
    }
    \hfill
    \addtocounter{subfigure}{-6}
    \subfloat[Original CT Image]{%
        \includegraphics[width=0.25\linewidth]{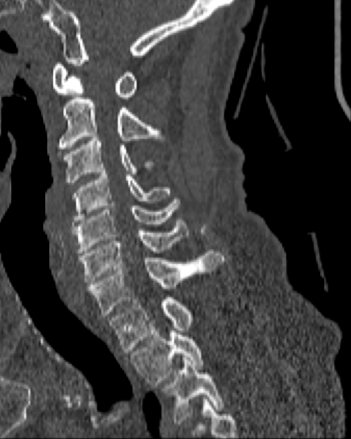}
        \label{fig:ct_image1}
    }
    \subfloat[Segmentation Result]{%
        \includegraphics[width=0.25\linewidth]{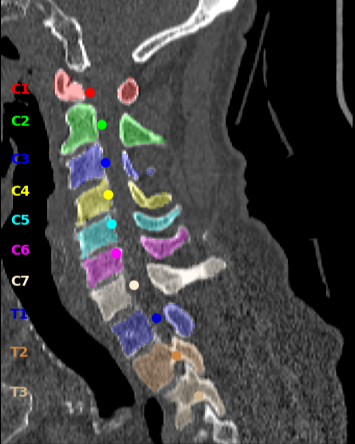}
        \label{fig:segmentation1}
    }
    \subfloat[3D Mesh]{%
        \includegraphics[width=0.25\linewidth]{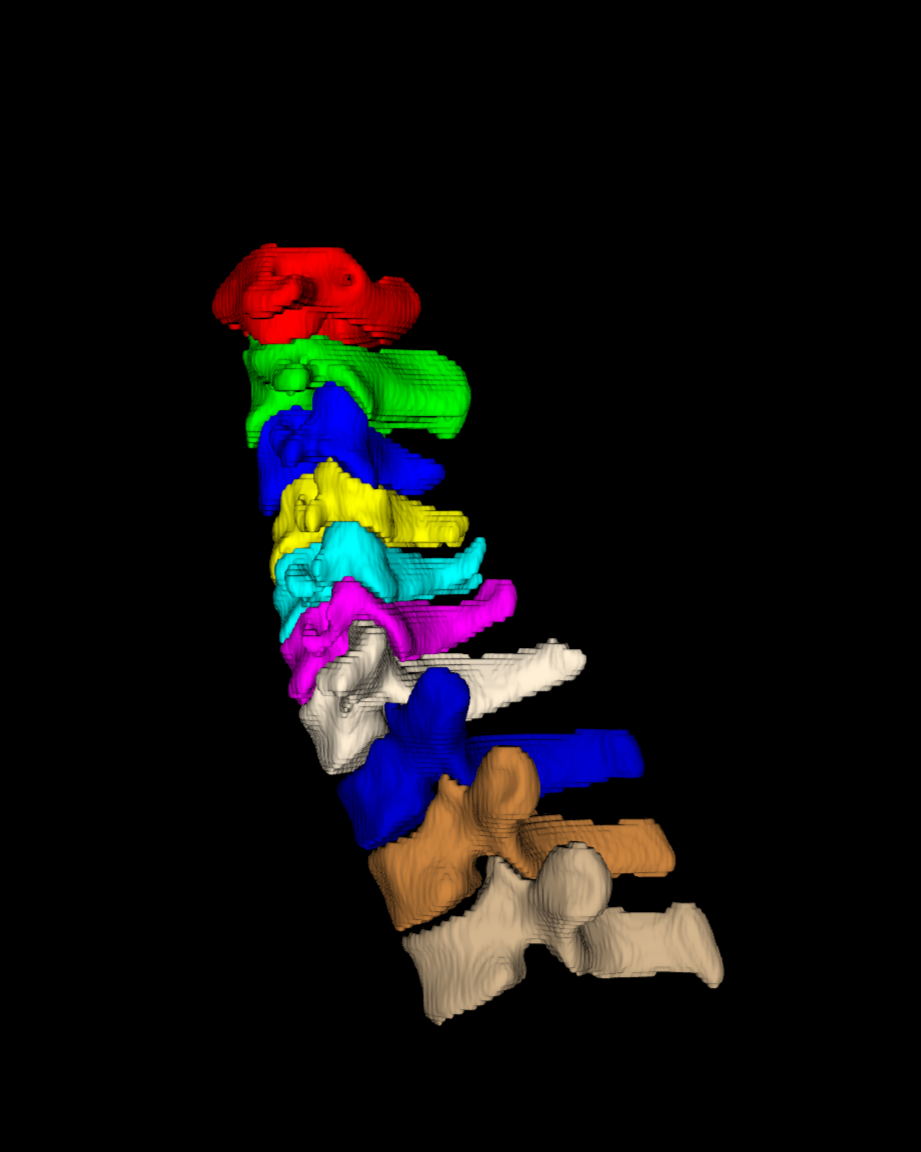}
        \label{fig:3d_mesh1}
    }
    \caption{Visualization of vertebral localization and segmentation results.}
    \label{fig:spine_visualization}
    \vspace{-0.75em}
\end{figure}

The models have 22.91 million parameters for spine localization, 21.13 million parameters for vertebrae localization, and 17.19 million parameters for vertebrae segmentation. The inference time varies with the field-of-view and the number of visible vertebrae in the CT volume. All experiments were conducted on a dual-socket Intel® Xeon® Platinum 8462Y+ system (64 physical cores, 2.80 GHz, 120 MiB L3 cache) running Linux. On the combined test set of 113 subjects, the average inference time per volume was about 1 min 40 s. Specifically, the process took about 3 s for spine localization, 19 s for vertebrae localization, and 1 min 17 s for vertebrae segmentation. The visualization of vertebral localization and segmentation is shown in \textbf{Figure~\ref{fig:spine_visualization}}.

\begin{figure}[htbp]
    \centering
    \subfloat[Missed vertebrae due to close centroids]{%
        \includegraphics[width=0.8\linewidth]{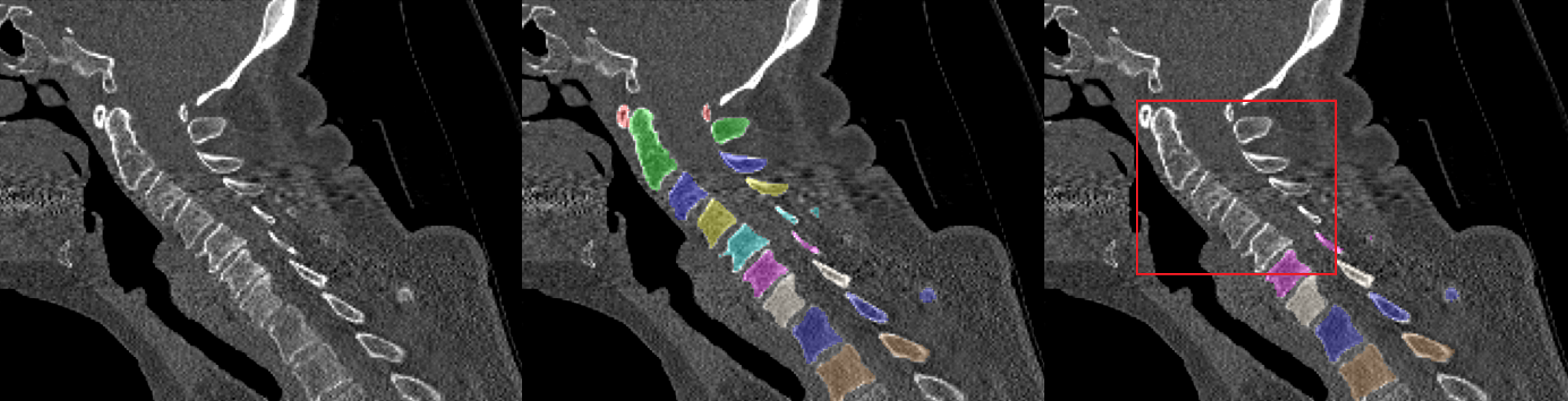}
        \label{fig:case1}
    }
    \hfill
    \subfloat[Label mismatch caused by vertebral mutation]{%
        \includegraphics[width=0.8\linewidth]{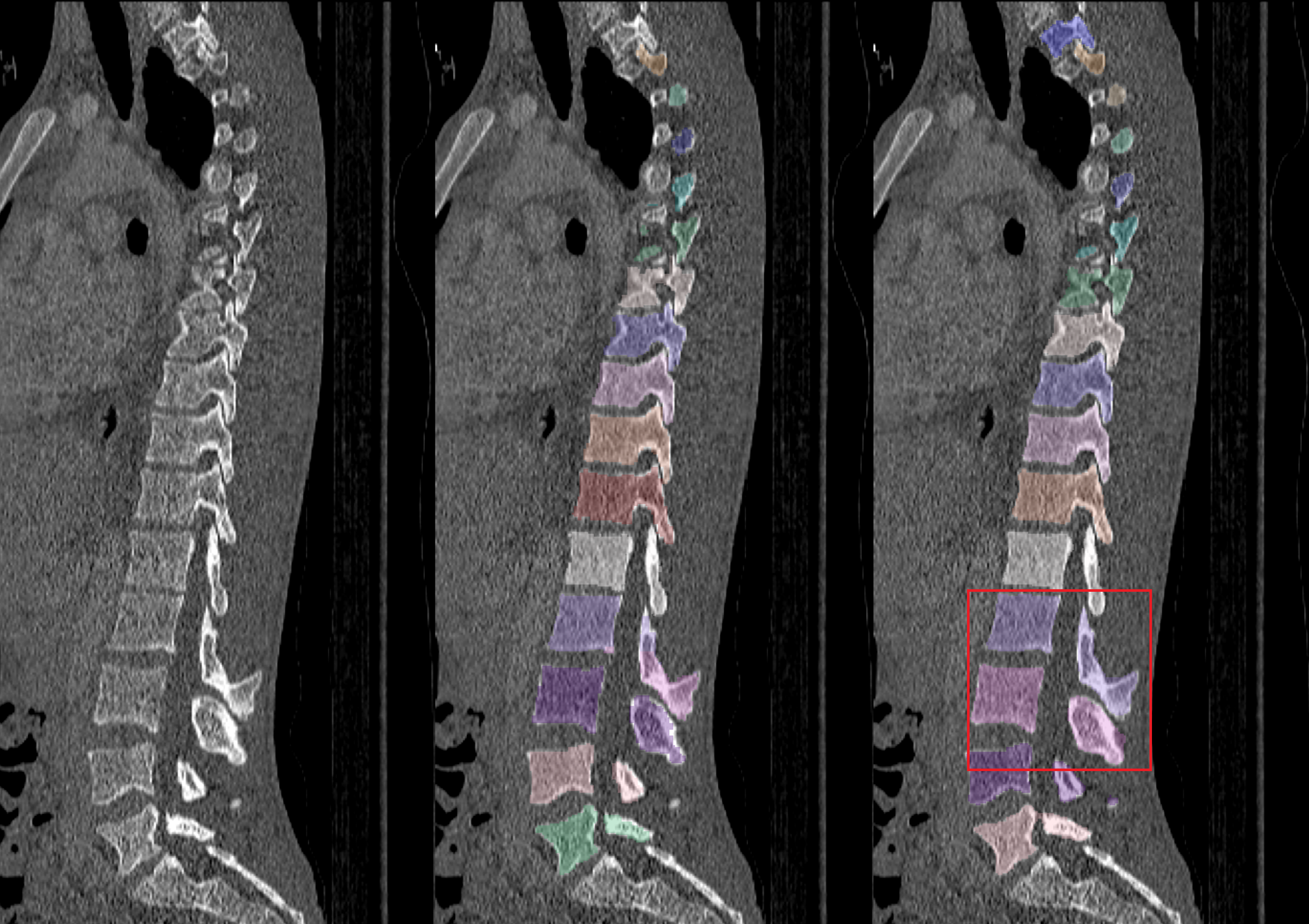}
        \label{fig:case2}
    }
    \hfill
    \subfloat[Label mismatch caused by metal implant artifacts]{%
        \includegraphics[width=0.8\linewidth]{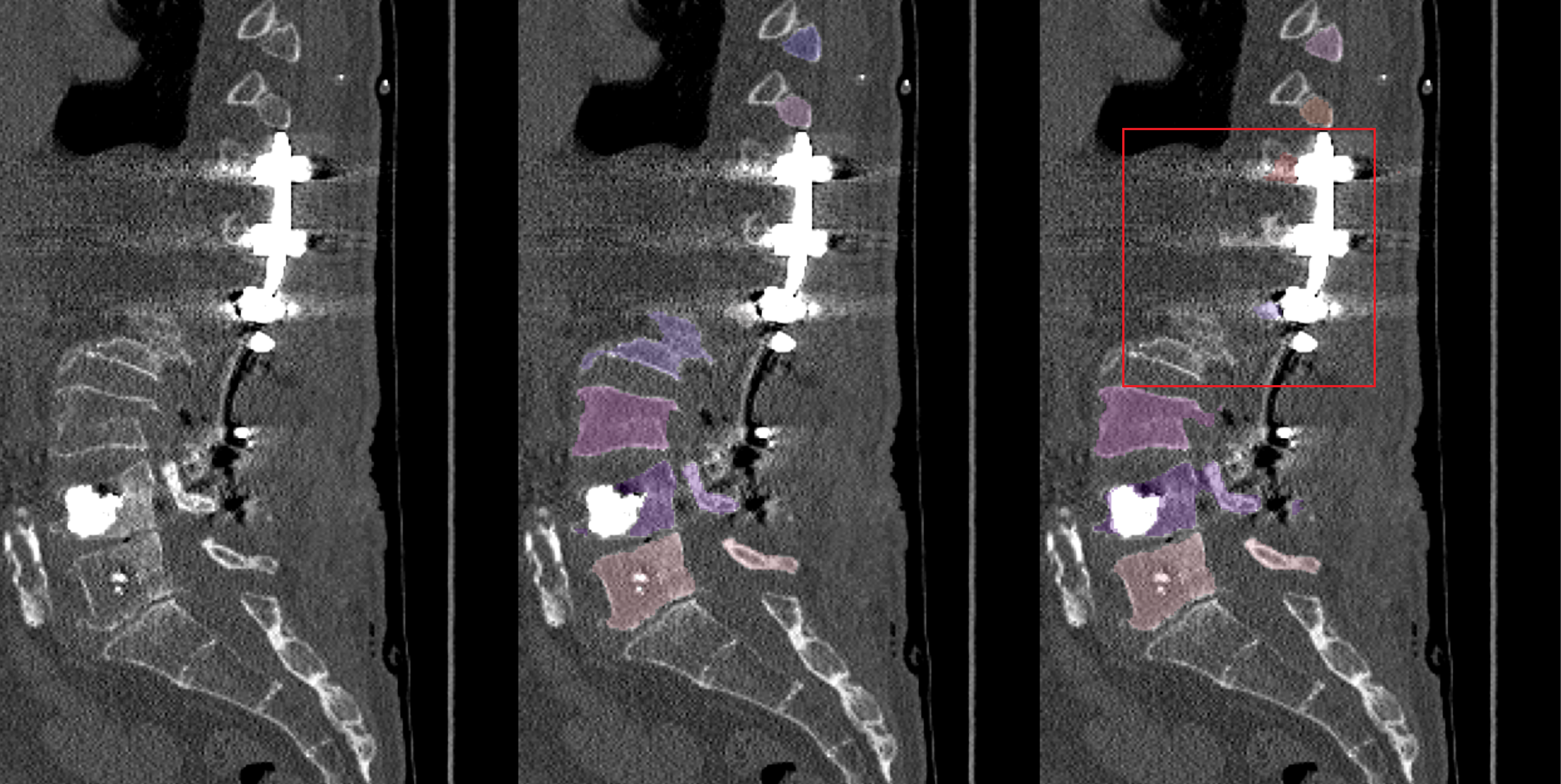}
        \label{fig:case3}
    }
    \caption{Visualization of representative failure cases. Each subfigure shows, from left to right, the original CT image, ground truth mask, and the predicted mask.}
    \label{fig:failure_cases}
    \vspace{-1.5em}
\end{figure}

Although the model demonstrated high performance overall, we observed several failure cases. In some cases like Figure~\ref{fig:case1}, vertebrae were missed completely. This is because vertebral centroids that are too close in z axis may cause blurred boundaries in segmentation, so these vertebrae centroids are deleted in the post-processing.  In some cases like Figure~\ref{fig:case2}, vertebrae were mismatched to incorrect labels due to rare vertebral mutation, or like in Figure~\ref{fig:case3}, due to artifact of metal implants. These issues highlight the challenges of accurate vertebra detection and segmentation in complex clinical scenarios.

%% file: 05-discussion.tex
\section{Discussions}
The EXR framework represents a foundational yet unvalidated prototype, warranting future usability and clinical validation studies to assess its impact on workflow efficiency, interpretability, and adoption. This work establishes the groundwork for advancing XR-based medical data visualization and AI integration. By unifying FHIR-standardized EHR data, multimodal imaging, and AI-driven analytics within an immersive environment, EXR introduces a paradigm shift in how clinicians, data engineers, and researchers interact with complex patient information.

Despite major advances in data standards and interoperability, conventional two-dimensional interfaces remain inherently limited in representing and conveying the spatial and temporal complexity of modern medical data. The presented proof-of-concept demonstrates that XR technologies can overcome these limitations by offering a unified, collaborative, and context-aware workspace for clinical decision-making.

Such systems have the potential to transform clinical workflows in several key ways. First, immersive environments can accelerate pattern recognition and enhance situational awareness by allowing clinicians to simultaneously explore patient histories, imaging studies, and AI-generated insights without switching between disparate systems. Second, XR-based platforms can foster multidisciplinary collaboration by enabling geographically distributed specialists to jointly visualize and manipulate patient-specific 3D workspaces in real time. Third, adherence to interoperability standards such as FHIR ensures both scalability and seamless integration with existing hospital infrastructures, while providing the flexibility to evolve alongside future healthcare data models.

The modular design of EXR also supports adaptation to dynamic clinical scenarios involving diverse conditions, temporal variations, and multimodal data sources. Future work will expand integration to additional FHIR resources—such as biosignals, clinical notes, and motion data—and evaluate EXR’s ability to visualize longitudinal disease trajectories, including progressive neurological and orthopedic conditions, across multiple encounters.

\subsection*{Future Work}
This work provides the foundations for future research and improvements. First, we aim to adopt a dynamical approach to the integration of AI pipelines, similarly to microservices, where the addition of an AI model and pipeline should lead to the minimum change in the main code. This would improve scalability to multiple models. Moreover, traditional data processing pipelines could be similarly integrated.
Second, we will develop a collaborative prototype to evaluate the different challenges of integrating multiple user in the same XR environment. We would first test local multiuser application, and if successful, explore the development of remote collaboration. Third, this work will lead to a validation and implementation research, notably with usability studies (using the System Usability Scale (SUS) and open-ended questionnaires), and finally adoption studies (usign for example the Technology Acceptance Model (TAM)) \cite{brooke1996sus, davis1989technology}.
Third, future work will focus on iteratively identifying potential universal representation of the different events (e.g., medications, observations, visits) intuitive for the users.
Lastly, as AI-driven analytics become embedded in XR environments, ethical and interpretability considerations must be prioritized. Future iterations of EXR will emphasize transparency and explainability through mechanisms such as confidence maps, saliency overlays, and textual annotations accompanying AI-derived outputs. In addition, intuitive patient-selection and summarization tools will be developed to highlight clinically relevant information, reduce cognitive load, and promote patient-centered visualization.

%% file: 06-conclusion.tex
\section{Conclusions}
This work introduces, to our knowledge, the first XR-based system integrating XR-based EHR multimodal data visualization integrating AI models.It demonstrates a proof-of-concept XR system that unifies FHIR-standardized EHR data with AI-driven medical imaging analysis in a collaborative, immersive environment. By integrating patient timelines, multimodal imaging, and automated 3D segmentation results, the platform enables clinicians, healthcare professionals, and researchers to interact with complex datasets to improve interpretation and context awareness, decision-making, and team collaboration. Our use-case highlights the potential of combining AI analytics with XR-based visualization to enhance understanding of anatomical structures and their clinical relevance. These results suggest that such integrated XR solutions could form the foundation for next-generation clinical decision-support tools, where advanced data infrastructures are directly accessible in an interactive and spatially rich environment.